\documentclass[groupedaddress,
 reprint,
 amsmath,amssymb,
 aps,
 prd,
]{revtex4-1}

\usepackage{graphicx}
\usepackage{dcolumn}
\usepackage{bm}
\usepackage{xcolor}
\usepackage{hyperref}
\usepackage[utf8]{inputenc}
\usepackage[capitalise]{cleveref}
\usepackage{commath}
\usepackage{booktabs}

\AtBeginDocument{
\heavyrulewidth=.08em
\lightrulewidth=.05em
\cmidrulewidth=.03em
\belowrulesep=.65ex
\belowbottomsep=0pt
\aboverulesep=.4ex
\abovetopsep=0pt
\cmidrulesep=\doublerulesep
\cmidrulekern=.5em
\defaultaddspace=.5em
}

\begin{document}

\title{Formation and mass growth of axion stars in axion miniclusters}
\author{Benedikt Eggemeier$^1$}
\author{Jens C. Niemeyer$^{1,2}$}

\affiliation{%
 $^1$Institut f\"ur Astrophysik, Universit\"at G\"ottingen, Germany\\
 $^2$Department of Physics, University of Auckland, Private Bag 92019, Auckland, New Zealand
}%

\date{\today}

\begin{abstract}
We study the formation and the subsequent mass growth of axion stars inside axion miniclusters. Numerically solving the Schr\"odinger-Poisson equations  
with realistic initial conditions we find that the axion stars exhibit similar properties to solitonic cores in ultralight bosonic dark matter halos in terms of their radial density profiles and large-amplitude oscillations. 
A merger of two axion stars confirms a previously found empirical law for the mass of the merged axion star. Monitoring the axion star masses over time, we observe a mass growth consistent with the mass increase of Bose stars in the kinetic regime reported by Levkov et al., confirming that the mass evolution of axion stars can be understood in terms of wave condensation. Based on this result, we predict a saturation of mass growth in relation to the minicluster mass consistent with the core-halo mass relation previously found for ultra-light bosonic dark matter halos.

\end{abstract}

\maketitle

\section{Introduction}
\label{sec:intro} 

Axion-like particles are very light pseudo-Nambu-Goldstone bosons of a spontaneously broken U(1) symmetry that generically couple very weakly to standard model fields, making them very attractive dark matter candidates. The original QCD axion, a by-product of the Peccei-Quinn (PQ) solution to the strong CP problem, is a particularly well motivated example \cite{PhysRevLett.40.223,PhysRevLett.40.279,PhysRevLett.43.103,shifman,DINE1981199,PRESKILL1983127,PQ1,PQ2}. Its theoretically preferred mass is a few times $10^{-5}$ eV \cite{Klaer:2017ond,Gorghetto:2018myk}. They are produced non-thermally by misalignment of the initial field value with very high occupation numbers and can be treated as a classical scalar field in the context of gravitational structure formation. In the case where the PQ symmetry is broken after inflation, spatially uncorrelated horizon-sized regions give rise to large isocurvature perturbations that can collapse into so-called axion miniclusters \cite{HOGAN1988228,PhysRevLett.71.3051,PhysRevD.49.5040,PhysRevD.50.769,Visinelli:minicluster}. 
The evolution of the axion field during the QCD phase transition and beyond is marked by the formation of strings and domain walls whose subsequent decay produces the seed inhomogeneities for miniclusters, as well as their gravitational collapse later during the radiation dominated epoch. As all of these processes are highly nonlinear, predicting the final abundance of cold axions and their clumping statistics relies on numerical simulations \cite{Vaquero2018,Buschmann:AxionSims}. With the axion mass left as the only relevant free parameter of the theory, these processes are responsible for the bulk of the theoretical uncertainties for forecasts and constraints from axion dark matter experiments.

The particular aspect we are concerned with in this work is the formation of axion stars inside axion miniclusters \cite{PhysRev.172.1331,PhysRev.187.1767,Tkachev_AS}. Axion stars are potentially observable by resonant decay, which can happen when axions are converted to photons in the magnetospheres of neutron stars~\cite{Tkachev:2014dpa,Pshirkov2009}, and have been suggested as a source for fast radio bursts~\cite{PhysRevLett.71.3051,Tkachev:2014dpa,Iwazaki:2014wka}. Besides, collisions with stellar objects have several observables, for example gravitational waves, neutrino emission and electromagnetic signals in a broad frequency range~\cite{Dietrich:2018jov,Raby:2016deh}. In the nonrelativistic limit relevant for cosmology, axion stars are bound states of the Schrödinger-Poisson (SP) equations that can form by classical Bose-Einstein condensation with purely gravitational interaction. The properties of axion stars, in particular their stability, are discussed in~\cite{VISINELLI_dense}. Furthermore, their formation has recently been investigated numerically by Levkov et al.~\cite{Levkov2018} in the kinetic regime where the axion field coherence length~$\sim (mv)^{-1}$ is much smaller than the characteristic scale of density variations (e.g. the minicluster radius $R$). We build upon their work using similar numerical techniques but different initial conditions. They were taken from lattice simulations of the axion field evolution throughout the QCD phase transition and hence consider the formation of strings and domain walls in the cosmological evolution of the axion field~ \cite{Vaquero2018}. With their large numerical simulations it is possible to analyze the small scale structure of the axion density field and its collapse into axion miniclusters. 

Our simulations robustly show the formation of axion stars in the core of miniclusters, confirming their existence under more realistic conditions with the caveat that an unrealistically low axion mass was used for numerical reasons. They form in a highly excited state with strong non-radial oscillations, as previously observed in the case of solitonic cores in ultralight axion dark matter halos \cite{Veltmaat:2018dfz}. We also confirm the initial mass growth of axion stars governed by the condensation time \cite{Levkov2018} and speculate about its eventual saturation when the virial velocity of the axion star exceeds that of the host minicluster. 

In the remainder of this work, we will describe our numerical methods and initial conditions in \cref{sec:methods}, present the simulation results in \cref{sec:results}, and discuss the consequences in \cref{sec:discussion}.

\section{Initial conditions and numerical methods}
\label{sec:methods}

The initial conditions for our simulations were taken from recent lattice simulations of the early axion field evolution~\cite{Vaquero2018}. The axion field was evolved as the phase of a complex scalar field with initial conditions uncorrelated over causally disconnected patches. A cosmic string network, due to the Kibble mechanism, is automatically included. As time evolves, the axion field increases its correlation length by smoothing inhomogeneities, collapsing loops and intersecting strings. The evolution continues until the axion mass becomes relevant, $H_1 = H(T_1) = m(T_1)$, when domain walls build up between strings leading to a fast destruction of the network. The rapid growth of the axion mass with temperature makes the axion field non-relativistic very fast, thus freezing the axion energy (dark matter) density distribution. At increasingly smaller scales $\sim 1/m$, long-lived oscillons (axitons) appear.  Although their number seems to increase with time, their size and relative importance appears to decrease. Axitons become unstable once the axion mass saturates (at the QCD confining temperature $T\sim 160$ MeV) and the small-scale axion field can free-stream efficiently. The current available grid resolutions forbid to resolve axiton cores until this time, but~\cite{Vaquero2018} showed that the large scale inhomogeneities (comoving wavenumber $k_1\lesssim 30 a_1H_1$) decouple from the small scale dynamics. Therefore, they switched off axion self-interactions before having resolution issues, allowing the axion field to free-stream away from axitons as it will do later on. This produced a smoothed axion dark matter distribution that is essentially frozen, which we use as initial conditions. The fast increase of the axion mass effectively suppresses the free-streaming of the high-density regions produced by the large tension of cosmic strings, the domain walls, and the first axitons and so the axion field exhibits large inhomogeneities even at the smallest resolved scales. The characteristic comoving length-scale of the simulations is the horizon size at $t_1$, which can be computed from the zero-temperature value of the axion mass $m=m(T=0)$~\cite{Vaquero2018}, 
\begin{align}
    \label{eq:L1}
    L_1 = \frac{1}{a_1 H_1} = 0.0362\left(\frac{50\,\mu\mathrm{eV}}{m}\right)^{0.167}\,\mathrm{pc}\,. 
\end{align}
The exact value of $m$ is not required as input as long as the temperature dependence of the axion mass is $m\sim 1/T^n$ with $n= 7$ (the value used in the simulations). Therefore one can use simulations for any value of $m$ above $\sim 10^{-9}$ eV. 
Ideally, $m$ will be fixed by the total dark matter yield, $\Omega_a h^2=\Omega_a h^2(m)=0.12$, if axions account for all the observed dark matter. However, the function $\Omega_a h^2(m)$ is subject to a relatively large uncertainty due to the need of extrapolating the simulations to realistic values of the string tension, see \cite{Gorghetto:2018myk} and references therein. A new method~\cite{Klaer:2017qhr} which produces effectively the correct tension predicts $m=26\pm 3\mu$eV~\cite{Klaer:2017ond} for the axion dark matter mass, while the direct method~\cite{Gorghetto:2018myk} currently has much larger errors $m\in(15\mu{\rm eV},\sim 10^3{\rm eV})$. Even in this generous range the value of $L_1$ varies at most by a factor of two. 

As initial conditions for this work we have chosen a simulation with boxsize $L = 6 \, L_1$ and $3096^3$ grid from~\cite{Vaquero2018} to simulate individual halos. The final density distribution was saved as a $512^3$ grid. 
We identified $206^3$-cell subvolumes with strong local overdensities in the original $512^3$ box. In order to enforce periodic boundary conditions demanded by our numerical scheme, each box was placed in the center of a new volume with a grid size of $256^3$ and we assigned the mean density of the subvolume to the 
boundary region with a width of 25 cells. A smooth transition from the subvolume to the boundary values was achieved by a smoothing procedure described below. Additionally, we interpolated the fiducial grid by a refinement factor of two onto a $512^3$ grid to improve the spatial resolution. 

We centered a Gaussian kernel multiplied with the corresponding density value on each fiducial cell. The smoothed density at the position $\mathbf{r_i}$ of a cell center on the refined grid is given by the sum of the overlapping density distributions at $\mathbf{r_i}$,
\begin{equation}
       \rho\,(\mathbf{r_i}) = \sum_{\mathbf{r'}}\rho(\mathbf{r'})W_G(\mathbf{r_i}-\mathbf{r'})\,H(|\mathbf{r_i}-\mathbf{r'}|)\,,
\end{equation}
where $\mathbf{r'}$ is the position of the cell centers of the fiducial grid, $W_G(\mathbf{r_i}-\mathbf{r'})$ is the Gaussian window function 
\begin{equation*}
    W_G(\mathbf{r_i}-\mathbf{r'}) = \frac{1}{(2\pi)^{3/2}\,\sigma^3}\exp\left(-\frac{(\mathbf{r_i}-\mathbf{r'})^2}{2\sigma^2}\right)\,,
\end{equation*}
and $H(|\mathbf{r_i}-\mathbf{r'}|)$ is a cut-off function defined by 
\begin{equation*}
    H(|\mathbf{r_i}-\mathbf{r'}|) = \begin{cases} 0\;, & |\mathbf{r_i}-\mathbf{r'}| \geq 5\,\sigma \\
     1\;, & |\mathbf{r_i}-\mathbf{r'}| < 5\,\sigma\end{cases}
\end{equation*}
with $\sigma=1.9$. 

In the non-relativistic approximation, a massive scalar field $\psi$ in comoving coordinates is described by the SP equations
\begin{align}
    \label{eq:SP}
    i\frac{\partial\psi}{\partial \tau} &= -\frac{1}{2}\nabla^2\psi + aV\psi \cr
    \nabla^2\psi &= \abs{\psi}^2 - 1 \,,
\end{align}
where $V$ is the Newtonian gravitational potential and $a$ is the cosmic scale factor. The comoving length is normalized such that $\boldsymbol{x}_{\text{comov}} = (\frac{3}{2}H_0^2\Omega_{m,0})^{1/4}(m/\hbar)^{1/2}\boldsymbol{x}$ and the comoving timestep is given by $d\tau = (\frac{3}{2}H_0^2\Omega_{m,0})^{1/2}a^{-2}dt$ with the present Hubble parameter $H_0$, the present dark matter density parameter $\Omega_{m,0}$ and the axion mass $m$. The comoving mass density is defined as $\rho = \abs{\psi}^2$ and normalized to the comoving mean density $\bar{\rho}$. The SP equations are solved using the fourth-order pseudo-spectral method described in \cite{Du2018}.
\begin{table*}[t]
    \caption{Axion star masses, together with comoving $r_\ast$, and corresponding halo masses for the three miniclusters at different redshifts. The two axion stars in the second minicluster MC2-AS1 and MC2-AS2 merged giving MC2-AS3. Shown also are the saturation mass from \cref{eq:McMh}, the condensation times obtained from the mass growth of the axion stars (cf. \cref{fig:mass_growth,eq:fit_condtime}) and the dimensionless parameter from \cref{eq:condtime}, where $b$ and $b_\mathrm{vir}$ are determined using the measured mean velocity and the virial velocity of the host minicluster, respectively.}
    \vspace{5pt}
    \begin{tabular}{@{}rccccccc@{}}
    \toprule
         & \multicolumn{1}{c}{$z$} & \multicolumn{1}{c}{$M_\ast\,[10^{-12} M_\odot]$} & \multicolumn{1}{c}{$r_\ast\,[10^{-3}\,\mathrm{pc}]$} &\multicolumn{1}{c}{$M_h\,[10^{-11} M_\odot]$} & \multicolumn{1}{c}{$M_{\ast,\mathrm{sat}}\,[10^{-12}\,M_\odot]$} &  \multicolumn{1}{c}{$\tau\,[10^7\,\mathrm{yr}]$} & \multicolumn{1}{c}{$b\,(b_\mathrm{vir})$}\\\midrule
         MC1-AS1 & $1277$ & $5.04$ & $1.36$ & $5.98$ & $5.96$ & $9.64$ & $0.11\,(0.61)$\\
         MC2-AS1 & $642$ & $3.17$ & $1.10$ & $16.6$ & $-$ & $-$ & $-$\\
         MC2-AS2 & $604$ & $2.65$ & $1.24$ & $5.53$ & $-$ & $-$ & $-$\\
         MC2-AS3 & $534$ & $3.83$ & $0.76$ & $17.7$ & $4.51$ & $15.1$ & $0.11\,(0.69)$\\
         MC3-AS1 & $899$ & $3.91$ & $1.74$ & $5.29$ & $4.74$ & $4.54$ & $0.10\,(0.96)$\\
         
    \bottomrule
    \end{tabular}
    \label{tab:MCs}
\end{table*}

The initial wavefunction $\psi$ is calculated from the smoothed and interpolated density field assuming a constant initial phase, corresponding to negligibly small initial velocities:
\begin{equation*}
    \text{Re}[\psi] = \sqrt{\rho} \quad,\quad \text{Im}[\psi] = 0\,.
\end{equation*}
In order to maintain adequate spatial resolution of the axion coherence length $\hbar/mv$ and the axion star radius throughout the simulation we chose an axion mass of $m = 10^{-8}$ eV and a box side length of $L = 0.356$ pc/$h$.

Furthermore, when a minicluster begins to form the density field is interpolated onto a $1024^3$ grid using the conservative second-order interpolation algorithm taken from \cite{Bryan:2013hfa}. 

We note that our choice for $m$ is lower by approximately three orders of magnitude than the current best fit to the dark matter abundance. Its role in our simulations is on the one hand to set the physical scales for the box size (which depends only weakly on $m$, cf. \cref{eq:L1}) and the axion star radius $r_\ast \sim m^{-2}$. On the other hand, the axion mass affects both the condensation time $\tau\sim m^3$ and the mass increase of the axion star (cf. \cref{eq:condtime,eq:fit_condtime}). Thus, the mass growth is less noticeable for higher axion masses.
As we are primarily concerned with the formation and early evolution of individual axion stars, leaving questions about the statistical distribution of their masses and densities for future work, we can justify our choice with the scaling symmetry of \cref{eq:SP} and the assumption that the structure of initial density perturbations depends only weakly on scale. Under these conditions, we argue that our main results are qualitatively valid as well for axion masses consistent with the dark matter abundance. 

All of our simulations start at redshift $z=7 \times 10^5$ using $\Omega_{m,0} = \Omega_{a,0} = 0.32$, $\Omega_{r,0} = 9.4\times 10^{-5}$, $\Omega_{\Lambda,0} =0.679906$, and $H_0=100 h$ km s$^{-1}$ Mpc$^{-1}$ with $h=0.67$.

\section{Simulation results}
\label{sec:results}

We present the simulations of three separate miniclusters (see \cref{tab:MCs}). 
\begin{figure}[b]
    \includegraphics[width=\columnwidth]{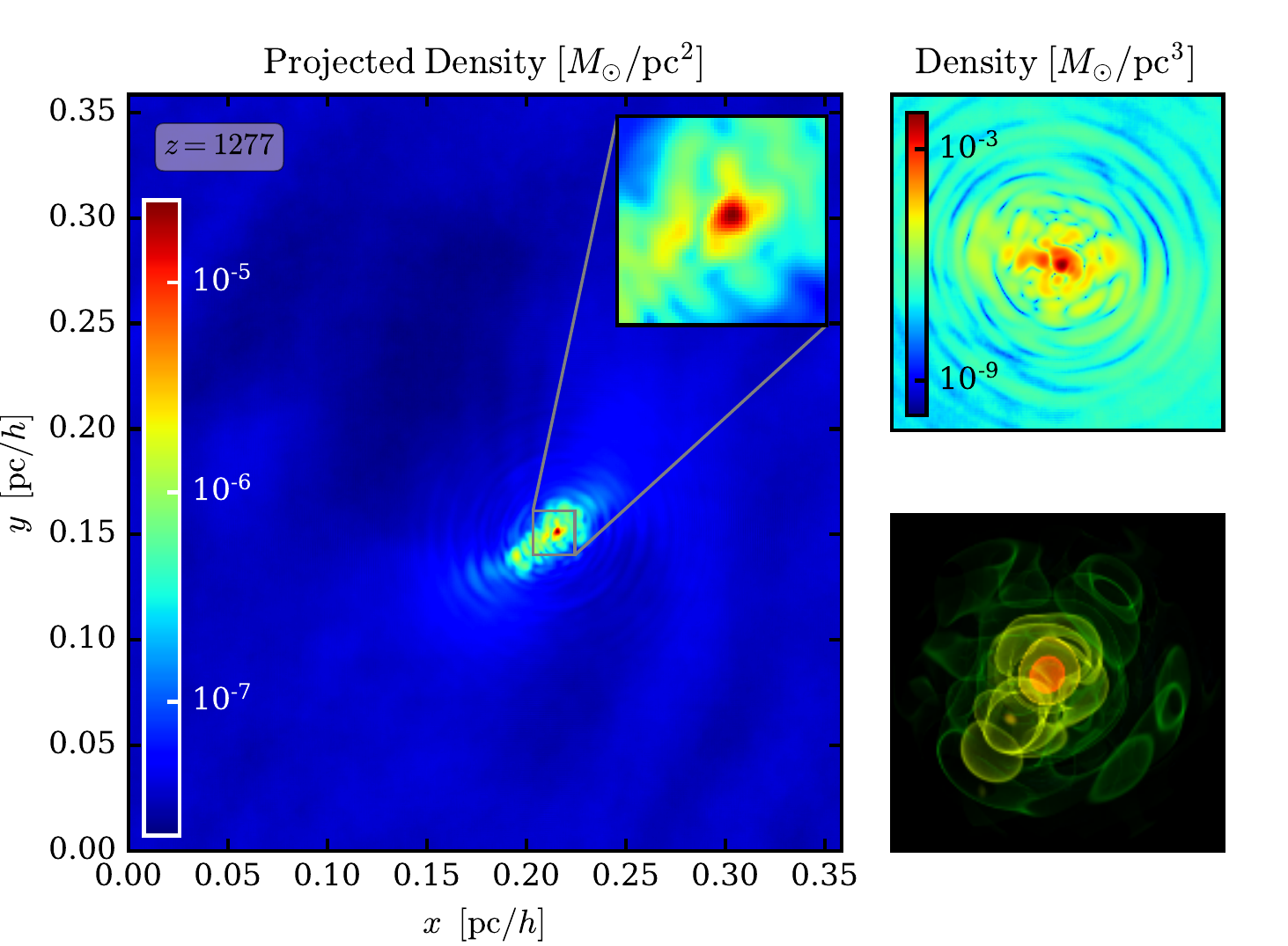}
    \caption{The left panel displays the projected density of a typical simulation in comoving units. The large box shows the full simulation domain. A zoom-in of the region where the axion star has formed is shown in the inlay. The granular structure within a radius of $0.021\,\mathrm{pc}/h$ can be more clearly seen in the slice plot in the upper right panel. A volume rendering of the axion star is shown in the lower right panel.}
    \label{fig:halo2_dens}
\end{figure}
One of the clusters consisted of two local density maxima that each produced an axion star which subsequently merged (see the discussion at the end of this section). Hence, we observed the evolution of five axion stars in total.

The overall structure of all miniclusters and their central axion stars is closely analogous to dark matter halos that form in fuzzy dark matter (FDM) simulations from cosmological initial conditions \cite{Schive:2014dra,Veltmaat:2018dfz}. Outside of the axion stars, the miniclusters consist of incoherent granular density fluctuations produced by wave interference. Additionally, the axion stars are surrounded by pronounced density waves. \Cref{fig:halo2_dens} shows a representative snapshot.

The axion stars form roughly during a few free-fall times after the collapse of the minicluster. Their exact formation time is ambiguous owing to the violent oscillations of proto-axion stars discussed below. 
\begin{figure}[t]
    \includegraphics[width=\columnwidth]{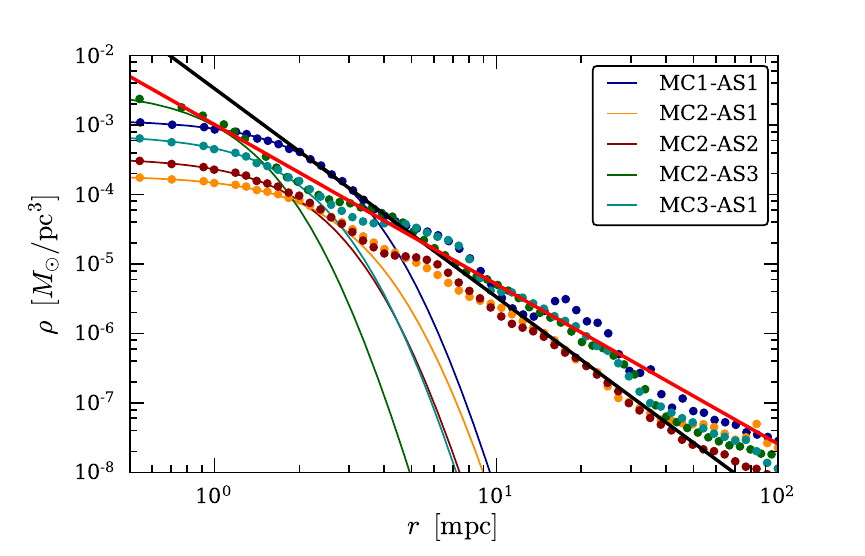}
    \caption{Density profiles of the five axion stars at different redshift (cf. \cref{tab:MCs}). The solid lines represent
    the theoretical profiles from \cref{eq:dens_profile} while the dots denote the data points. The black solid line corresponds to $r^{-3}$ as expected for the outer parts of an NFW profile and the red solid line to $r^{-9/4}$ for the outer profile.}
    \label{fig:densprofiles}
\end{figure}
We define an axion star by the existence of a self-bound, cored central density whose angle-averaged profile is well described by an approximate Bose star solution (e.g. \cite{Schive:2014hza}):
\begin{align}
    &\rho_\ast(r) \simeq \rho_0 \left[1+0.091 \left(\frac{r}{r_\ast}\right)^2 \right]^{-8}\,, \cr 
    &\rho_0 = 1.9\times 10^{-6}\,a^{-1} \cr 
    &\qquad \times \left(\frac{10^{-8}\,\mathrm{eV}}{m}\right)^2  \,\left(\frac{10^{-3} \mathrm{pc}}{r_\ast}\right)^4\; \frac{M_\odot}{\mathrm{pc}^3}\,,
    \label{eq:dens_profile}
\end{align}
where $\rho_0$ is the comoving central core density and $r_\ast$ is defined by the comoving radius at which the density drops to half of its maximum value. By calculating the virial parameter $|E_\mathrm{pot}|/(2\times E_\mathrm{kin})$, we verified that the axion stars are in virial equilibrium. The radial density profiles of the five axion stars and their corresponding theoretical profiles are shown in \cref{fig:densprofiles}. It can also be seen that the outer density profiles of the incoherent halo decline less steeply than the asymptotic NFW power-law of $r^{-3}$. Their slope is consistent with $r^{-9/4}$ predicted for isolated perturbations accreting from a homogeneous background~\cite{Bertschinger1985}.

\begin{figure}[t]
    \includegraphics[width=\columnwidth]{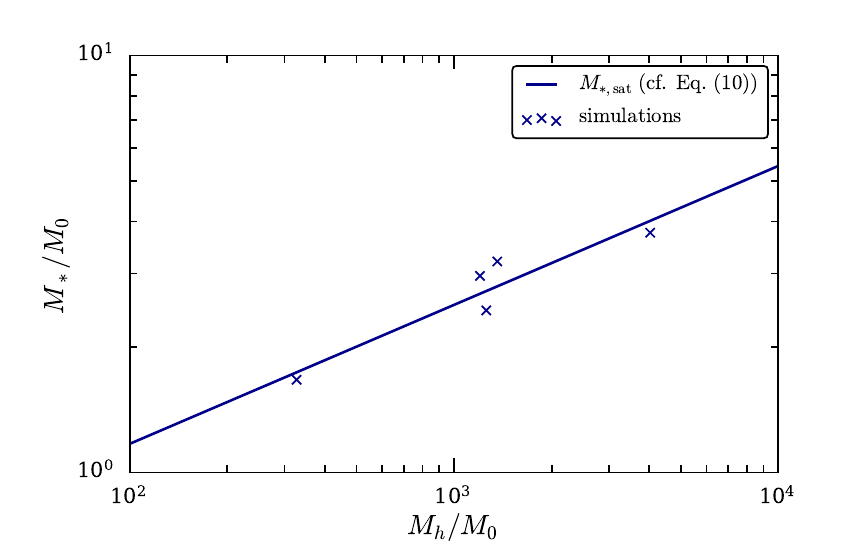}
    \caption{Axion star masses as a function of their host halo masses, determined at the time at which a stable axion star has formed. The solid line shows the prediction from saturated mass growth in \cref{eq:McMh} consistent with the core-halo mass relation for FDM halos found in \cite{Schive:2014hza} with $M_0\sim 4.4\times 10^{-14}(m/10^{-8}\,\mathrm{eV})^{-3/2}\,M_\odot$.}
    \label{fig:corehalo}
\end{figure}

We define the axion star mass $M_\ast$ as the mass inside a sphere with (physical) radius $r_\ast$ \cite{Schive:2014hza}:
\begin{equation}
M_\ast = 5.4\times 10^{-15}\left(\frac{10^{-8}\, \mathrm{eV}}{m} \right)^2 \left(\frac{10^{-3} \mathrm{pc}}{r_\ast} \right) \, M_{\odot}\,.
\end{equation}
In \cref{fig:corehalo}, $M_\ast$  is plotted for all five axion stars against the masses of their host halos. The solid line shows the prediction for the axion star mass at the point of saturated mass growth from \cref{eq:McMh}; see \cref{sec:discussion} for a detailed discussion. It obeys the scaling relation $M_\ast \sim M_h^{1/3}$ also found for solitonic cores in FDM halos \cite{Schive:2014hza}.

\begin{figure}
    \includegraphics[width=\columnwidth]{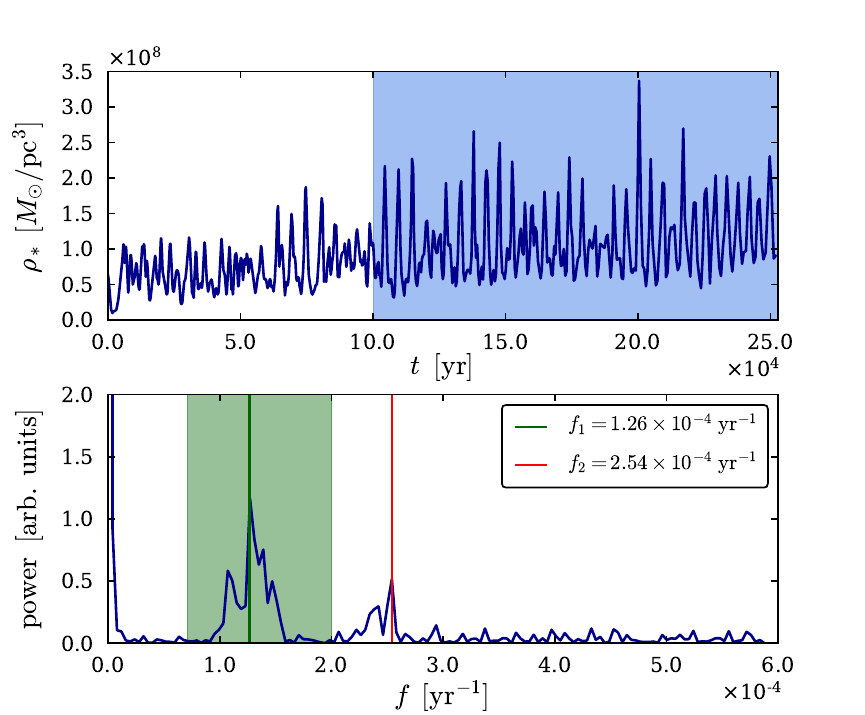}
    \caption{Oscillation of the axion star MC1-AS1. In the upper panel, one can see the (physical) axion star density fluctuations. The blue colored region marks the data which was used for the Fourier transformation which is shown in the lower panel. Before $t=10^5\,\mathrm{yr}$ the axion star is still forming, which is why this was not considered in the Fourier transformation. The boundaries of the green colored region are the expected quasi-normal frequencies (cf. Eq.~(\ref{eq:osc_freq})) for the maximum and the minimum of the axion star density in the blue colored region, respectively. The green line represents the frequency peak at the quasi-normal frequency $f_1=1.26\times 10^{-4}\,\mathrm{yr}^{-1}$ and the red line the second dominating frequency peak at $f_2 = 2.54\times 10^{-4}\,\mathrm{yr}^{-1}$.}
    \label{fig:halo1_osc}
\end{figure}

Analyzing the evolving density field with high temporal resolution reveals that the axion stars oscillate with amplitudes of more than a factor of two and density-dependent frequencies. \Cref{fig:halo1_osc} shows the fluctuating axion star density and its temporal Fourier transform for the axion star MC1-AS1. The blue shaded region marks the data used for the Fourier transformation. The frequency spectrum has distinctive peaks at the first two quasi-normal modes,
\begin{equation}
    f_1 = 1.1\times 10^{-4}\left(\frac{\rho_\ast}{10^8\,M_{\odot}\text{pc}^{-3}}\right)^{1/2}\,\mathrm{yr}^{-1}\,,
    \label{eq:osc_freq}
\end{equation}
and   
$f_2\approx 2 f_1$ \cite{guzman,Guzman:2018bmo}. Using the mean value of $\rho_\ast$ in the blue colored region in the upper panel of \cref{fig:halo1_osc} yields $f_1 = 1.13\times 10^{-4}$ yr$^{-1}$. 

\begin{figure}
    \centering
    \includegraphics[width=\columnwidth]{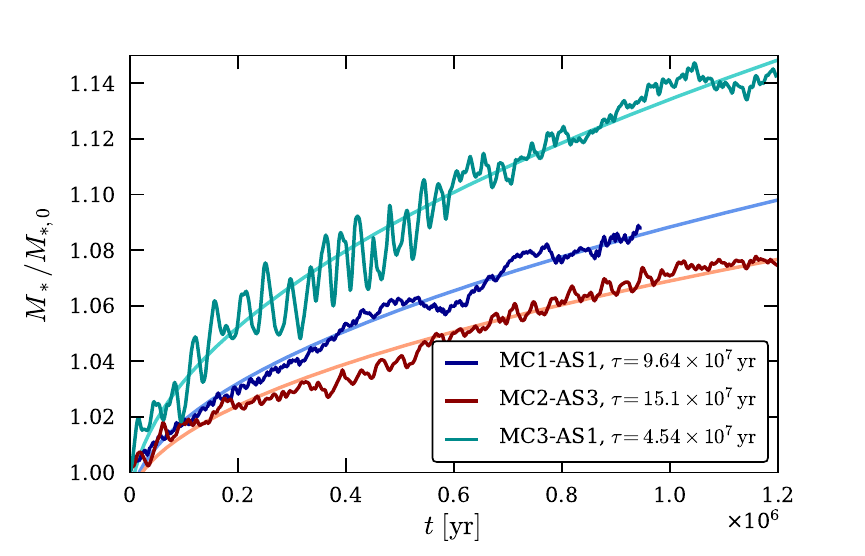}
    \caption{Evolution of the masses of three axion stars relative to their initial mass, evaluated at the earliest time the soliton profile \cref{eq:dens_profile} provided a good fit, compared to \cref{eq:fit_condtime}.}
    \label{fig:mass_growth}
\end{figure}

\cref{fig:mass_growth} displays the masses of three axion stars as a function of time, normalized by the masses at a reference time when the soliton profile began to be a good fit to the radial density profile.  
The axion star masses increase by about $15\%$ over a period of $\sim 1.2 \times 10^6$ yr, with a time dependence consistent with the mass growth of Bose stars in the kinetic regime observed in \cite{Levkov2018}:
\begin{equation}
    M_\ast(t)\simeq M_{\ast,0}\left(\frac{t}{\tau}\right)^{1/2}\,.
    \label{eq:fit_condtime}
\end{equation}
The condensation time $\tau$ is given in terms of the radius $R$, velocity $v$ and density $\rho$ of the halo as 
\begin{equation}
    \tau \simeq \frac{\sqrt{2}b}{12 \pi^3} \left(\frac{\hbar}{m}\right)^{-3} \frac{v^6}{G^2 \rho^2 \log \Lambda}\,,\,\Lambda \sim \frac{R}{\hbar/m v} \,,
    \label{eq:condtime}
\end{equation}
with a coefficient $b$ that needs to be computed numerically. The results of fitting \cref{eq:fit_condtime} to the growth curves in \cref{fig:mass_growth} can be found in \cref{tab:MCs}. For determining $b$, we used $v = \langle \vert \nabla S/m \vert \rangle$, where the average was taken over the virial radius $R_h$ of the minicluster, yielding  $b\simeq 0.1$. As the measured values of $v$ are systematically higher than the virial velocities  $v_\mathrm{vir}$ of the miniclusters by $\sim 30-50\%$, the $v^6$-dependence of $\tau$ results in significantly higher values for $b_\mathrm{vir}$ defined in terms of $v_\mathrm{vir}$ (also shown in \cref{tab:MCs}).

As mentioned at the beginning of this section, we observed a merger of two axion stars in one of our simulations. 
We resolved two progenitor stars (MC2-AS1, MC2-AS2) and the merged axion star (MC2-AS3) as can be seen in \cref{fig:densprofiles}. Taking the masses from \cref{tab:MCs}, we find that the mass of the merged axion star is $M_{\ast,3} = \beta (M_{\ast,1}+M_{\ast,2})$ with $\beta \sim 0.66$, while $\sim 34\%$ of the total initial mass was radiated away by gravitational cooling. This is consistent with simulations of binary mergers of boson stars giving  $\beta \sim 0.7$ \cite{Schwabe:2016rze}. The additional mass loss in the present case may be a consequence of the highly excited state of the progenitor axion stars.

\section{Discussion}
\label{sec:discussion}

Our simulations, in particular the confirmation of the time-dependent mass growth of the axion stars obeying \cref{eq:fit_condtime,eq:condtime} observed by Levkov et al. \cite{Levkov2018}, provide further evidence that the mass evolution of axion stars in miniclusters can be explained in terms of a kinetic process of wave condensation \cite{TKACHEV1991289} (we will comment about the formation itself below). We can use this framework to predict the mass at which the growth of the axion star saturates by forming a local cloud of field fluctuations whose temperature exceeds the virial temperature of the host halo. This relation turns out to coincide exactly with the core-halo mass relation found empirically for FDM solitonic cores \cite{Schive:2014hza}.

Immediately after an axion star has formed, the state of its ambient axion field is governed by the virial temperature of the minicluster, i.e. $v \simeq v_\mathrm{vir,mc}$ in \cref{eq:condtime}. After it has grown to a critical mass, the axion star forms a hot axion atmosphere with the star's virial temperature, at which point $\tau$ itself becomes dependent on $M_\ast$. This causes the mass growth to saturate and slow down substantially. 

The transition takes place when $v_\mathrm{vir,mc} \simeq v_\mathrm{vir,\ast}$ where the virial velocity in the gravitational potential of the axion star is approximately given by \citep{Hui2017}
\begin{equation}
    \label{eq:vvir_as}
    v_\mathrm{vir,\ast}(M_\ast) \simeq \frac{G M_\ast m}{\hbar} \,.
\end{equation}
Approximating the minicluster as a uniform sphere with (physical) radius $R_h$ and mass $M_h = (4 \pi/3) R_h^3 \zeta(z) \rho_{m,0}/a^3$ (with $\zeta(z\gg1) \simeq 18 \pi^2$), its virial velocity is $v_\mathrm{vir,mc}^2 = 3 G M_h/ 10 R_h$. The saturation criterion is therefore met when 
\begin{equation}
    \label{eq:McMh}
    M_{\ast,\mathrm{sat}} = \left(\frac{\hbar}{m}\right)\, 
    \left(\frac{3 }{10\,a\,G}\right)^{1/2}\, \left(\frac{4 \pi\, \zeta(z)\, \rho_{m,0}}{3}\right)^{1/6}\,M_h^{1/3}
\end{equation}
($ M_\ast \leq M_h$). 

\Cref{eq:McMh} coincides exactly with the relation found in \cite{Schive:2014hza} for the final axion star mass. However, ongoing condensation predicts that the axion star mass continues to grow, albeit at a drastically reduced rate. Inserting $v_\mathrm{vir,\ast}(M_\ast)$ into \cref{eq:condtime} and assuming that the power law growth continues to hold, the axion star mass will eventually follow
\begin{equation}
    M_\ast(t) \simeq M_{\ast,\mathrm{sat}}\,\left(\frac{t}{\tau_\mathrm{sat}}\right)^{1/8}\,,
    \label{eq:secular}
\end{equation}
where $\tau_\mathrm{sat}$ follows from evaluating \cref{eq:condtime} with $v=v_\mathrm{vir,\ast}(M_{\ast,\mathrm{sat}})$. For many purposes, the axion star mass can therefore be estimated by \cref{eq:McMh} but secular growth may be important in some cases. Simulations will have to verify if the long-term mass growth asymptotically approaches \cref{eq:secular}.

We finish with some remarks on the validity of the wave condensation formalism. The kinetic regime, in which the Wigner distribution $f_W$ for $\psi$ has been shown to obey a kinetic equation sourced by the Landau scattering term $\sim f_W/\tau$ with $\tau$ from \cref{eq:condtime} in \cite{Levkov2018}, is valid if $\hbar (mv)^{-1} \ll R$ where $R$ is the characteristic scale of the minicluster. Although the power-law density profiles of our miniclusters are scale-free, the observed mass growth controlled by $\tau$ indicates that the kinetic description holds after virialization of the minicluster. 
However, the kinetic description is not valid throughout our full simulation. Starting from the non-kinetic case, the gravitational collapse of the initial overdensities and the subsequent formation of virialized halos lead to a period in which the kinetic regime is finally entered. During the first few free-fall times, it comes to a phase of violent relaxation where strong density fluctuations occur on all scales and the background gravitational potential is time-dependent.
The \emph{formation} of axion stars prior to complete virialization hence cannot be explained unambiguously by wave condensation (this also applies to solitonic cores in FDM halos \cite{Veltmaat:2018dfz}). It is therefore not too surprising that the axion stars in our simulation form significantly earlier than predicted by their condensation time. We conjecture that the violent relaxation phase further enhances the probability to form axion stars particularly near the center of axion miniclusters, making the existence of at least one star per cluster much more likely than suggested by the condensation time.

\section{Conclusions}
\label{sec:conclusions}

Using a pseudo-spectral method to solve the Schrödinger-Poisson (SP) equations we studied the formation and evolution of axion stars in the center of axion miniclusters from realistic initial conditions. We confirm that the density profiles of the axion stars are in accordance with ground-state solutions of the SP equations and those of solitonic cores in FDM halos~\cite{Schive:2014dra} while the outer density profiles of miniclusters are close to $r^{-9/4}$ as predicted in~\cite{Bertschinger1985}. We do not address the statistical distribution of masses and density profiles here, leaving these questions to future work. 

Monitoring the mass of the axion stars over a period of $\sim 1.2\times 10^6$ yr, we found a mass growth by about $15\%$ consistent with the mass increase of Bose stars in the kinetic regime observed in~\cite{Levkov2018}. Thus, we confirm that the mass evolution of axion stars in axion miniclusters can be explained in terms of a kinetic process of wave condensation. We predict a  decreasing mass growth as $t^{1/8}$ once the virial velocity of the axion star and the axion minicluster coincide. The corresponding saturation mass is exactly the one found in~\cite{Schive:2014hza}, providing a dynamical explanation for their result. 

We observed that axion stars form in highly excited states with strong quasi-normal oscillations with amplitudes of more than a factor of two. Hence, we  confirm \cite{Veltmaat:2018dfz} which used a different numerical method that might have been affected by noise from the boundaries. This is not the case in our simulations showing that the oscillations have a physical origin. 

A merger of two axion stars supports the empirical law for the mass of the merged axion star found in~\cite{Schwabe:2016rze}. In contrast to their simulations, the merging axion stars in our scenario are in highly excited states explaining a higher mass loss due to gravitational cooling compared to~\cite{Schwabe:2016rze}. 

Overall, our work provides further evidence that the existence of axion stars is a firm prediction of scenarios in which dark matter consists of QCD axions and the Peccei-Quinn symmetry is broken after inflation, with important consequences for potential astrophysical observations.

\acknowledgements{This work would have been impossible without Javier Redondo's help and the initial conditions provided by him, Alejandro Vaquero, and Julia Stadler. We thank them first and foremost, together with Xiaolong Du for assistance with his pseudo-spectral code. We would also like to thank Richard Easther, Mateja Gosenca, Shaun Hotchkiss, Emily Kendall, Dmitry Levkov, Doddy Marsh, Alexander Panin, Bodo Schwabe, Igor Tkachev, and Jan Veltmaat for helpful comments and discussions. We acknowledge the \textsc{Yt} toolkit~\cite{yt} that was used for our analysis of numerical data. JCN acknowledges funding by a Julius von Haast Fellowship Award provided by the New Zealand Ministry of Business, Innovation and Employment and administered by the Royal Society of New Zealand.}

\bibliography{AS_formation}

\end{document}